# Electrical characterization of gadolinium oxide deposited by high pressure sputtering with in situ plasma oxidation

María Ángela Pampillón, Pedro Carlos Feijoo, Enrique San Andrés

Dpto. Física Aplicada III, Universidad Complutense de Madrid, Avda. Complutense, S/N, 28040 Madrid, Spain

**Abstract**

In this work, we characterized gadolinium oxide films deposited on silicon by high pressure sputtering with a two-step process: first, we sputtered metallic gadolinium in an argon atmosphere and then, we performed an in-situ plasma oxidation of the metallic layer previously deposited. By means of high resolution transmission electron microscopy, we can detect the oxidation degree of the metallic film. Under optimized deposition conditions, fully oxidized $Gd_2O_3$ films are obtained. In addition, the capacitance and conductance as a function of gate voltage of Pt gated metal–insulator–semiconductor capacitors confirm stable dielectric behavior of the fully oxidized films. The devices show low gate leakage currents ($10^5$ A/cm$^2$ at 1 V for 2.2 nm of equivalent oxide thickness), low interface trap density and an almost negligible hysteresis and frequency dispersion.

## 1. Introduction

The future downscaling of MOSFETs requires the use of alternative high permittivity (high k) dielectrics beyond hafnium oxide based materials [1] due to their low crystallization temperature (around 500 C) [2]. Currently, for this purpose, there are many dielectrics under investigation. Among them, gadolinium oxide presents interesting properties, including a k value near 15, a band gap around 6 eV and chemical stability with Si [3,4] and with III–V semiconductors [5–7]. Despite the moderate permittivity, this dielectric is interesting since the combination of $Gd_2O_3$ with $Sc_2O_3$ produces gadolinium scandate with a reported k value around 23 [8,9]. This value is much higher than the permittivity of HfSiO(N), used in CMOS advanced devices. One of the challenges of these dielectrics is to minimize the regrowth of a $SiO_x$ interfacial layer between the high k material and Si after the dielectric deposition. This undesirable low k film would limit the equivalent oxide thickness (EOT) of the structure and, therefore, the effective permittivity value of the stack. In this work we studied $GdO_x$ obtained from a metallic Gd deposition by high pressure sputtering (HPS) followed by an in situ plasma oxidation with the aim of minimizing the interface $SiO_x$ regrowth [10]. By using HPS we want to minimize plasma damage of the Si surface, since at the working pressure of 50 Pa, the sputtered species get thermalized in a distance of less than 1 mm. We studied several Gd deposition and oxidizing plasma conditions to evaluate the electrical and structural properties of the resulting films to fully oxidize the Gd layer while minimizing the $SiO_x$ regrowth.

## 2. Experimental

Gadolinium oxide films were grown on n-type 2 inch Si wafers with (100) orientation and resistivity of 1.5–10.0 Ω cm. Prior to the high k deposition, 200 nm of field oxide ($SiO_2$) were e-beam evaporated and a positive lithographic process was used in order to open square windows with sizes ranging from 10 × 10 to 700 × 700 um$^2$. The aim of this step is to insulate the devices and build up the pads (the device top contact). The two-step method used for $GdO_x$ deposition by HPS is described in detail in Refs. [10,11]. Two series of samples with different Gd deposition times (80, 120 and 160 s) and plasma oxidation lengths (100 and 300 s) were studied. The top electrode of the MIS devices was deposited by e-beam evaporation of 50 nm of Pt or a 4 nm Pt/50 nm Al stack. Pt is a noble metal and would not react with the dielectric films. The backside contact was a 50 nm Ti/100 nm Al stack. A 20 min forming gas annealing (FGA) was performed at temperatures between 300 and 400 C. These devices were electrically characterized before and after the FGA. The gate capacitance and conductance as a function of gate voltage characteristics were measured at frequencies from 1 kHz to 1 MHz. Gate leakage currents, hysteresis and hard breakdown voltages were also measured. High resolution transmission electron microscopy (HRTEM) images of the MIS capacitors were obtained and also the energy dispersive X-ray spectroscopy (EDX) when using the microscope in the scanning TEM (STEM) mode. X-ray photoelectron spectroscopy (XPS) was used to obtain surface dielectric film composition.

### 3. Results and discussions

The cross-sectional HRTEM image of Fig. 1 evidences that the $GdO_x$ with 80 s of Gd deposition and 300 s of plasma oxidation is amorphous, with a thickness of 6 nm and flat interfaces. However, a 1.5 nm $SiO_x$ interlayer appears between the $Gd_2O_3$ and the Si substrate due to an excessive oxidation length. According to XPS analysis (not shown here) [12], the top surface of the gadolinium oxide film has an O/Gd ratio of $1.52 \pm 0.05$, which means that this plasma oxidation process produces stoichiometric $Gd_2O_3$. Fig. 2 shows an HRTEM image of a device with a Gd film twice as thick (the metal was deposited for 160 s) also oxidized during 300 s. There, we observe a stacked structure, with a darker layer of 5 nm on top of a lighter one 4 nm thick. Both layers are amorphous. EDX measurements showed the Gd and O presence on both layers. The contrast observed on Fig. 2 suggests different oxidation degrees of the $GdO_x$ film. The upper film would be fully oxidized $Gd_2O_3$ while the lower one would be a Gd suboxide. This could be due to an oxidation process not long enough. The oxygen introduced into de chamber performs the oxidation of the Gd film, starting from the upper layers and then, diffusing deeper. If the oxidation time is too short, the complete oxidation is not achieved. But if this oxidation length is excessive, a $SiO_x$ layer appears, as it can be seen in Fig. 1. This would explain the fact that doubling the metal deposition time does not yield a layer twice as thick. Furthermore, in Fig. 2 there is no evidence of a $SiO_x$ interface layer regrowth, supporting the incomplete oxidation explanation. In order to confirm this hypothesis, it would be necessary an in-depth analysis by XPS or high resolution electron energy loss spectroscopy with the aim to obtain quantitative information on the depth composition of these films.

Since the TEM image of the thinner film showed that a plasma oxidation of 300 s was excessive, we fabricated MIS devices with metal deposition times of 80 s and 120 s, but with a shorter plasma oxidation of only 100 s. Figs. 3 and 4 present the $C_{gate}$ and g of these samples before and after the FGA. For the thinner samples (in Fig. 3) there is not a great difference between capacitance values before and after FGA, indicating a stable $Gd_2O_3$. The EOT of these samples is 2.2 nm, which assuming a 6.0 nm thick $Gd_2O_3$ film would yield a $k_{eff}$ of about 11. This value is calculated by supposing that there is no $SiO_x$ interface. If a thin 0.5 nm $SiO_x$ interface were used in the calculation, the plasma oxidized $Gd_2O_3$ permittivity would be 14, very close to the reported value of stoichiometric $Gd_2O_3$, around 15 [3,13]. For the thicker samples (with 120 s of Gd deposition), a capacitance drop and flatband voltage instability with annealing temperature are observed (Fig. 4). The capacitance drop can be explained by a $SiO_x$ regrowth at the $GdO_x$/Si interface.

From the EOT change from 2.9 nm before annealing to 3.6 nm after FGA at 400 C, we can estimate a 0.7 nm $SiO_x$ regrowth. The flatband variations are due to charge trapping of detrapping of bulk traps, so these changes suggest a high amount of traps, possibly due to dangling bonds in the $GdO_x$/Si interface. Then, these electrical results also supports the previous conclusion that an oxidation time too short produces a substoichiometric and unstable $GdO_x$ in contact with the Si substrate. Using the conductance method, we obtained that the density of states ($D_{it}$) is reduced one order of magnitude after the FGA for the thinner sample, from $10^{12}$ eV$^1$cm$^2$ to $2 \cdot 10^{11}$ eV$^1$cm$^2$ after annealing at 400 C. On the other hand, the $D_{it}$ reduction is smaller for the thicker films: from $6 \cdot 10^{11}$ eV$^1$cm$^2$ to $2 \cdot 10^{11}$ eV$^1$cm$^2$ after annealing. For both types of samples, the values obtained after the FGA at 400 C are low enough to ensure a reasonable MOSFET operation.

Fig. 5 represents the gate leakage current as a function of gate voltage. As it can be seen, these gate currents are low: $10^3$ A/cm$^2$ for the thinner sample and below $10^4$ A/cm$^2$ for the thicker one, even at 3 V. These values are similar to those reported in other works [14]. Ref. [15] shows the J–V$_{gate}$ curves for different values of the SiO$_2$ thickness. There, it can be observed that for a sample 2.2 nm thick, the interpolated leakage current at 3 V is around 1 A/cm$^2$, three orders of magnitude higher than the current obtained in Fig. 5. In both cases, for the different FGA temperatures, the curves are similar and do not change significantly with the different annealing temperatures (only on the thinner devices there seems to be a consistent reduction of the leakage after annealing at 300 C value, possibly due to a slight interfacial regrowth and/ or Gd$_2$O$_3$ densification). Besides, in order to check the uniformity of the samples and also to obtain information on the maximum voltage applicable, in Fig. 6 we show the J–V$_{gate}$ curves of several devices for the thinner sample (80 s of Gd) annealed at 400 C, measured beyond the hard breakdown voltages. We observed that between 3.1 and 3.3 V the devices presented an increase of the slope of the leakage current (soft breakdown events). At voltages between 3.5 and 3.7 V all devices showed hard breakdown. These voltages are high enough for device applications, and the repeatability indicates a uniform and controlled Gd$_2$O$_3$ film.

Fig. 7 presents the C$_{gate}$–V$_{gate}$ hysteresis curves for both types of samples. The sweep started from accumulation to inversion and back again. These curves show a hysteresis of 80 mV for 80 s Gd sample (Fig. 7a) and an almost negligible value for the 120 s Gd sample (15 mV) (Fig. 7b). These small values are lower than those reported in other works [16]. This small charge trapping suggested that either the trap density is low or that the traps remain charged or discharged during the voltage sweep.

Finally, the fully oxidized devices (those with thinner Gd deposition time) present almost no frequency dispersion of the C$_{gate}$–V$_{gate}$ curves, as it is represented in Fig. 8. This structure only shows a small drop in the capacitance value at 1 MHz, due to the coupled effect of the high conductance and the series resistance. Besides, from this figure and using the conductance method, it can be obtained that the $D_{it}$ decreases slightly as the measurement frequency is increased, from $5 \cdot 10^{11}\,\text{eV}^1\,\text{cm}^2$ at 1 kHz to $3 \cdot 10^{11}\,\text{eV}^1\,\text{cm}^2$ at 1 MHz. This means that, most interface traps are able to respond even at moderately high frequencies.

## 4. Conclusion

In this work we have obtained amorphous $GdO_x$ layers by sputtering a Gd target in an Ar plasma and then, performing an in-situ plasma oxidation in $Ar/O_2$ atmosphere. We have observed that depending on the oxidation length, we obtained fully oxidized $Gd_2O_3$ layers (amorphous and stable) or two different oxidation degrees films (also amorphous but unstable). For the fully oxidized devices, the conductance and gate leakage remain low, even after several FGA at different temperatures. The hard breakdown voltage is above 3.5 V. An almost negligible hysteresis and frequency dispersion are also observed.

These results demonstrate that high pressure sputtering combined with plasma oxidation is a promising technique to obtain high k stacks with good electrical properties. At present we are adapting this process to obtain $Gd_xSc_{1x}O_3$, aiming at a higher $k_{eff}$. Acknowledgments

The authors would like to thanks 'C.A.I de Técnicas Físicas' of the 'Universidad Complutense de Madrid' and 'Instituto de Nanociencia de Aragón' of the 'Universidad de Zaragoza'. This work was funded by the project TEC2010-18051 and FPI program (BES2011-043798) of the Spanish 'Ministerio de Economía y Competitividad'.

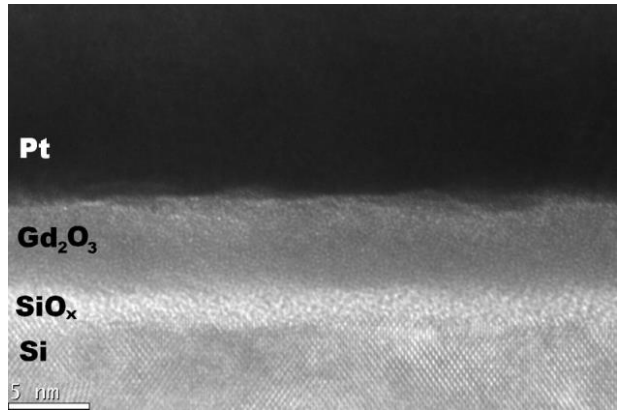

Fig. 1. Cross-sectional HRTEM image of a MIS device with FGA at 300 C and Pt gated electrode. GdO$_x$ was obtained after 80 s metal deposition and a 300 s plasma oxidation.

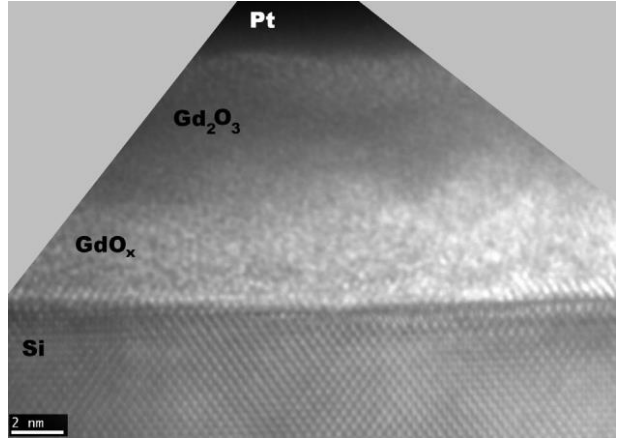

Fig. 2. Cross-sectional HRTEM image of a MIS device with FGA at 300 C and a Pt gated electrode. GdO$_x$ was obtained after 160 s of metal and 300 s of oxidation.

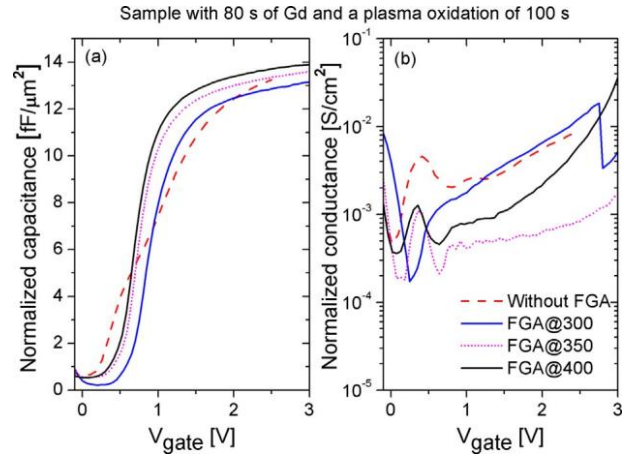

Fig. 3. (a) $C_{gate}$–$V_{gate}$ and (b) $G$–$V_{gate}$ curves at 10 kHz before and after FGA at different temperatures for sample with 80 s of Gd and a 100 s plasma oxidation.

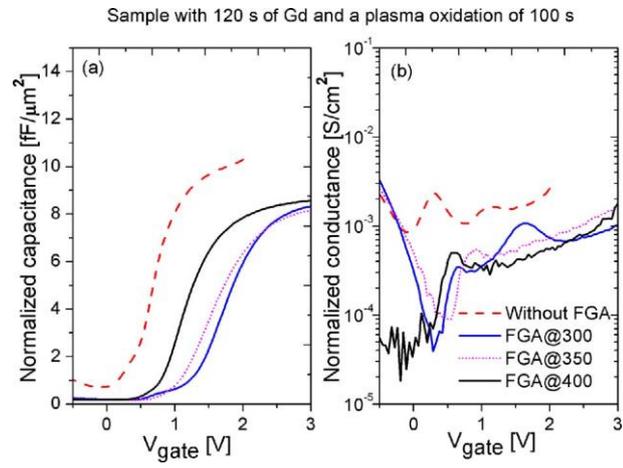

Fig. 4. (a) $C_{gate}$–$V_{gate}$ and (b) G–$V_{gate}$ curves at 10 kHz before and after FGA at different temperatures for sample with 120 s of Gd and a 100 s plasma oxidation.

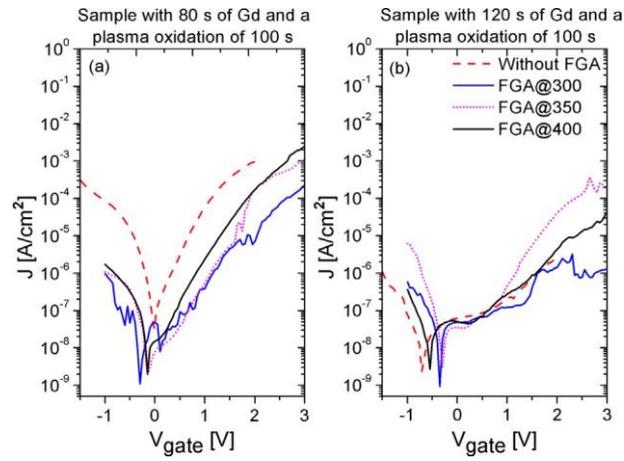

Fig. 5. J–V$_{gate}$ characteristics before and after FGA at different temperatures for (a) 80 s deposited Gd and (b) 120 s deposited Gd, both with a 100 s plasma oxidation.

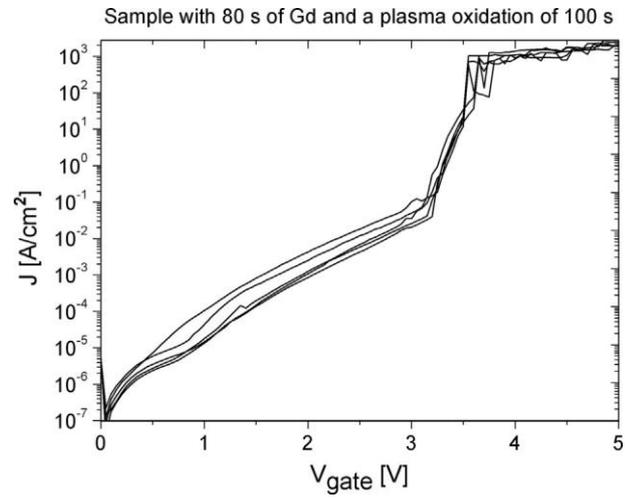

Fig. 6. J–V$_{gate}$ characteristics for the sample with 80 s of Gd and a 100 s oxidation after FGA at 400 C, measured in several devices to observe breakdown events and reproducibility.

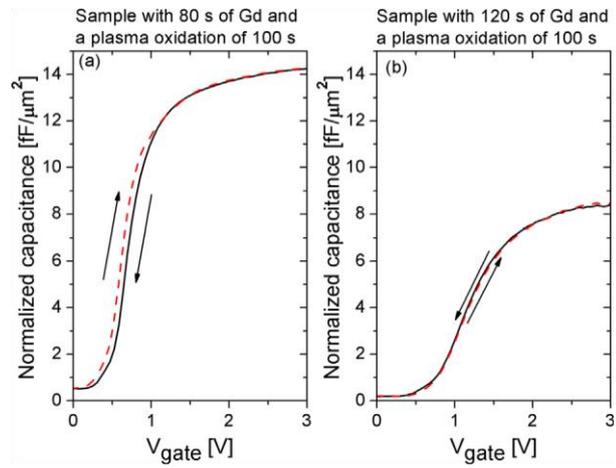

Fig. 7. $C_{gate}$–$V_{gate}$ hysteresis curves for (a) 80 s of Gd sample and (b) 120 s of Gd with 100 s of plasma oxidation at 10 kHz after FGA at 400 C.

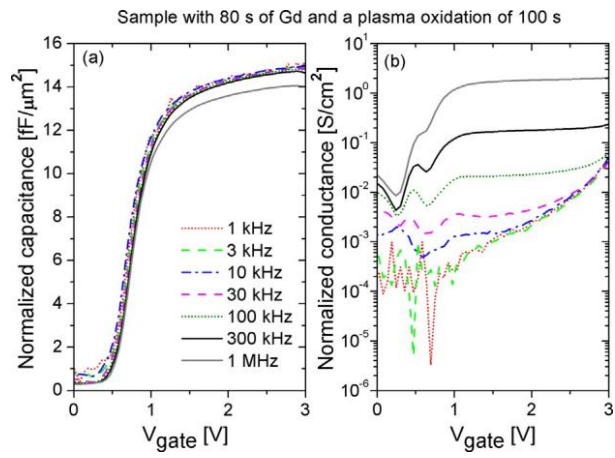

Fig. 8. (a) $C_{gate}$–$V_{gate}$ and (b) G–$V_{gate}$ curves at different frequencies after FGA at 400 C for sample with 80 s of Gd and 100 s of oxidation.